\documentclass[preprint,showpacs,preprintnumbers,amsmath,amssymb,pre]{revtex4}
\usepackage[english]{babel}
\usepackage[dvipdf]{graphicx}
\usepackage[dvips]{epsfig}
\usepackage{amssymb}
\usepackage{amsmath}
\usepackage{amsbsy}
\usepackage{dcolumn}

\usepackage[section]{placeins}
\usepackage{color}

\definecolor {myc} {rgb} {0,0,0}  

\begin{document}
\title{Rearrangement zone around a crack tip in a double self-assembled transient  network}

\author{Guillaume Foyart$^{1}$, Christian Ligoure$^{1}$, Serge Mora$^{2}$}
\author{Laurence Ramos$^{1}$}
\email{laurence.ramos@umontpellier.fr}

\affiliation{
$^1$ Laboratoire Charles Coulomb UMR 5221,
 CNRS, Universit\'{e} de Montpellier,
 F-34095, Montpellier, France\\
 $^2$ Laboratoire de Mécanique et de Génie Civil, UMR 5508,
Universit\'{e} de Montpellier and CNRS. 163 Rue Auguste Broussonnet. F-34090 Montpellier, France\\
}

\date{\today}

\begin{abstract}
We investigate the nucleation and propagation of cracks in self-assembled viscoelastic fluids, which are made of surfactant micelles reversibly linked by telechelic polymers. The morphology of the micelles can be continuously tuned, from spherical to rod-like to wormlike, thus producing transient double networks when the micelles are sufficiently long and entangled, and transient single networks otherwise. For a single network, we show that cracks nucleate when the sample deformation rate involved is comparable to the relaxation time scale of the network. For a double network, by contrast, significant rearrangements of the micelles occur as a crack nucleates and propagates. We show that birefringence develops at the crack tip over a finite length, $\xi$, which corresponds to the length scale over which micelles alignment occur. We find that $\xi$ is larger for slower cracks, suggesting an increase of ductility.
\end{abstract}

\maketitle

Despite fluid-like features, viscoelastic fluids display fracture processes with intriguing analogies with those of hard materials~\cite{Berret2001,Gladden2007,Skrzeszewska2010,Ligoure2013}. Hele-Shaw experiments based on the injection of a low viscosity fluid into the viscoelastic material confined between two plates has been proven as a simple, yet efficient and robust, way to induce and observe cracks in complex fluids. Such set-up has been successfully used to fracture granular media~\cite{Holtzman2012} and soft materials, including foams~\cite{Arif2010}, surfactant phases~\cite{Greffier1998, Yamamoto2006}, solutions of associating polymers~\cite{Zhao1993,Ignes1995,Vlad1999,Mora2010,Foyart2013}, and colloidal suspensions~\cite{Lemaire1991}. In most cases, a rule of thumb is that the material will fracture when solicited at sufficiently high rates so that dissipative processes do not have time to dominate the sample response \cite{Gladden2007,Tabuteau2009,Tripathi2006, Lindner2008}. For a single transient network characterized by a unique relaxation time, this simple rule implies that the rate involved is larger than the inverse of the relaxation time, avoiding dissipative processes and rendering the material brittle. When samples are by contrast characterized by several distinct relaxation times, dissipative processes might take place during crack nucleation and propagation, conferring therefore some ductility to the fracture process.

Here we use viscoelastic suspensions of surfactant micelles reversibly linked by triblock polymers. The morphology of the micelles can be continuously tuned, from spherical to wormlike, thus producing double networks when the micelles are sufficiently long and entangled, and single networks otherwise~\cite{Tixier2010}. In addition to the stress relaxation due to the transient un-bridging of the micelles, the anisotropic morphology of the wormlike micelles provides an additional mean for stress relaxation due to their alignment. This extra degree of freedom confers ductility to networks otherwise brittle (when the micelles are spherical)~\cite{Ramos2011}. We use a Hele-Shaw set-up to image the nucleation and propagation of cracks in those networks. We show that for double networks birefringence due to micelle alignment develops at the crack tip over a finite length scale, $\xi$, hinting at a link with the process zone in ductile materials. We find that $\xi$ increases as the velocity of the crack decreases, revealing an increase of the sample ductility.

\begin{figure}
\includegraphics[width=0.5\textwidth]{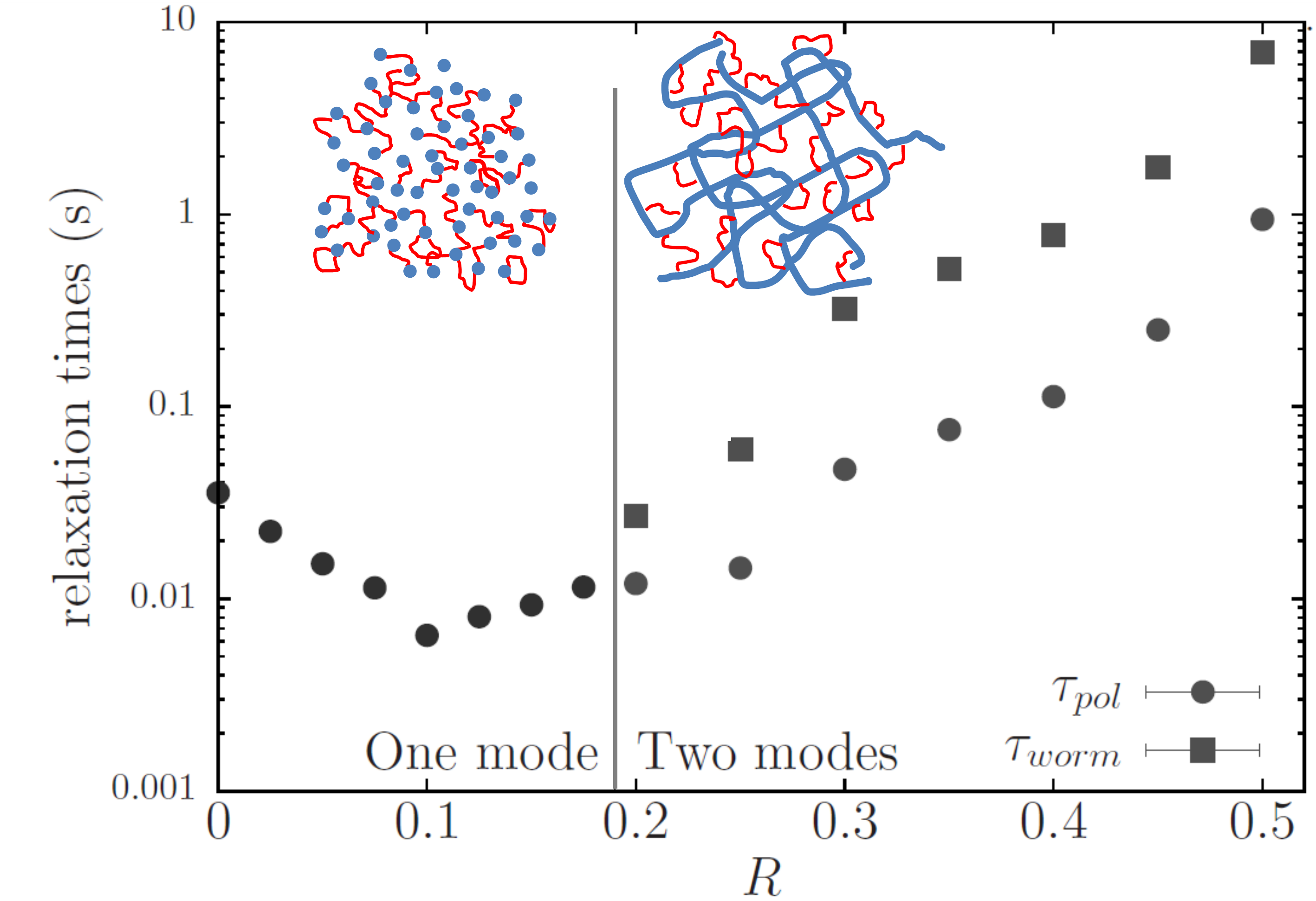}
\caption{Relaxation times of single and double transient networks. Characteristic relaxation times, as extracted from a fit of the frequency dependence of the complex modulus with a one-mode (resp. two-mode) Maxwell fluid model, for $R<0.2$ (resp. $R\geq0.2$), as a function of the growth factor $R$. Inset: Sketches of a single and a double networks. }
\label{fig:Rheo}
\end{figure}

The self-assembled transient networks comprise surfactant micelles dispersed in brine ($0.5$ M NaCl) and reversibly bridged by telechelic polymers. Micelles are composed of a mixture of cetylpyridinium chloride (CpCl) and sodium salicylate (NaSal) with a NaSal/CpCl molar ratio $R$ in the range $[0-0.5]$. As shown previously~\cite{Tixier2010,Ramos2011} the micelles continuously elongate as $R$ increases, allowing their morphology to be tuned from spherical micelles to rods to very long and entangled (wormlike) micelles (inset fig.~\ref{fig:Rheo}). Telechelic polymers are made of a long water-soluble polyethylene oxide chain (molecular weight $10$ kg/mol), flanked at each extremity by a short carboxylic chain with $23$ carbons.
The mass fraction of the micelles $\varphi=(m_{\rm{CpCl}} + m_{\rm{NaSal}})/m_{\rm{total}}$, resp. the amount of polymer $\beta = m_{\rm{polymer}}/(m_{\rm{CpCl}} + m_{\rm{NaSal}})$, is set at $9$ \%, resp. $55$ \%. Here $m_{\rm{CpCl}}$, $m_{\rm{NaSal}}$, and $m_{\rm{polymer}}$ are respectively the mass of CpCl, NaSal and polymer, and $m_{\rm{total}}$ is the total mass of the sample. Samples made of spherical micelles and short rod micelles behave as Maxwell fluids~\cite{Tixier2010}. They are characterized by one elastic modulus and one relaxation time $\tau_{\rm{pol}}$. $\tau_{\rm{pol}}$ is related to the bridging dynamics of the telechelic polymers, i.e. to the time scale for the disengagement of the hydrophobic stickers of the polymer from the core of the micelles and their subsequent re-anchoring. Above a critical $R$ ($R^* \approx 0.2$), the micelles become so long that they entangle and form themselves a viscoelastic network. At this stage, the samples behave as two-mode Maxwell fluids~\cite{Nakaya2008,Tixier2010}, resulting from the coexistence of two coupled networks, one related to the bridging of the micelles by the telechelic polymer (relaxation time $\tau_{\rm{pol}}$) and one related to the micelle entanglement (relaxation time $\tau_{\rm{worm}}$). $\tau_{\rm{worm}}$ is related to the dynamics of disentanglements due to the combination of reptation and micelle breaking and recombination~\cite{Nakaya2008,Cates1987}. Figure~\ref{fig:Rheo} displays the evolution with the growth factor $R$ of the rheological relaxation times, $\tau_{\rm{pol}}$ and $\tau_{\rm{worm}}$. The characteristic time related to the polymer network varies non monotonically with $R$: $\tau_{\rm{pol}}$ decreases up to $R=0.1$ (in the one-mode Maxwell fluid regime) and then increases. $\tau_{\rm{worm}}$ on the other hand is systematically larger than $\tau_{\rm{pol}}$ and increases steadily with $R>R^*$.

\begin{figure}
\includegraphics[width=0.5\textwidth]{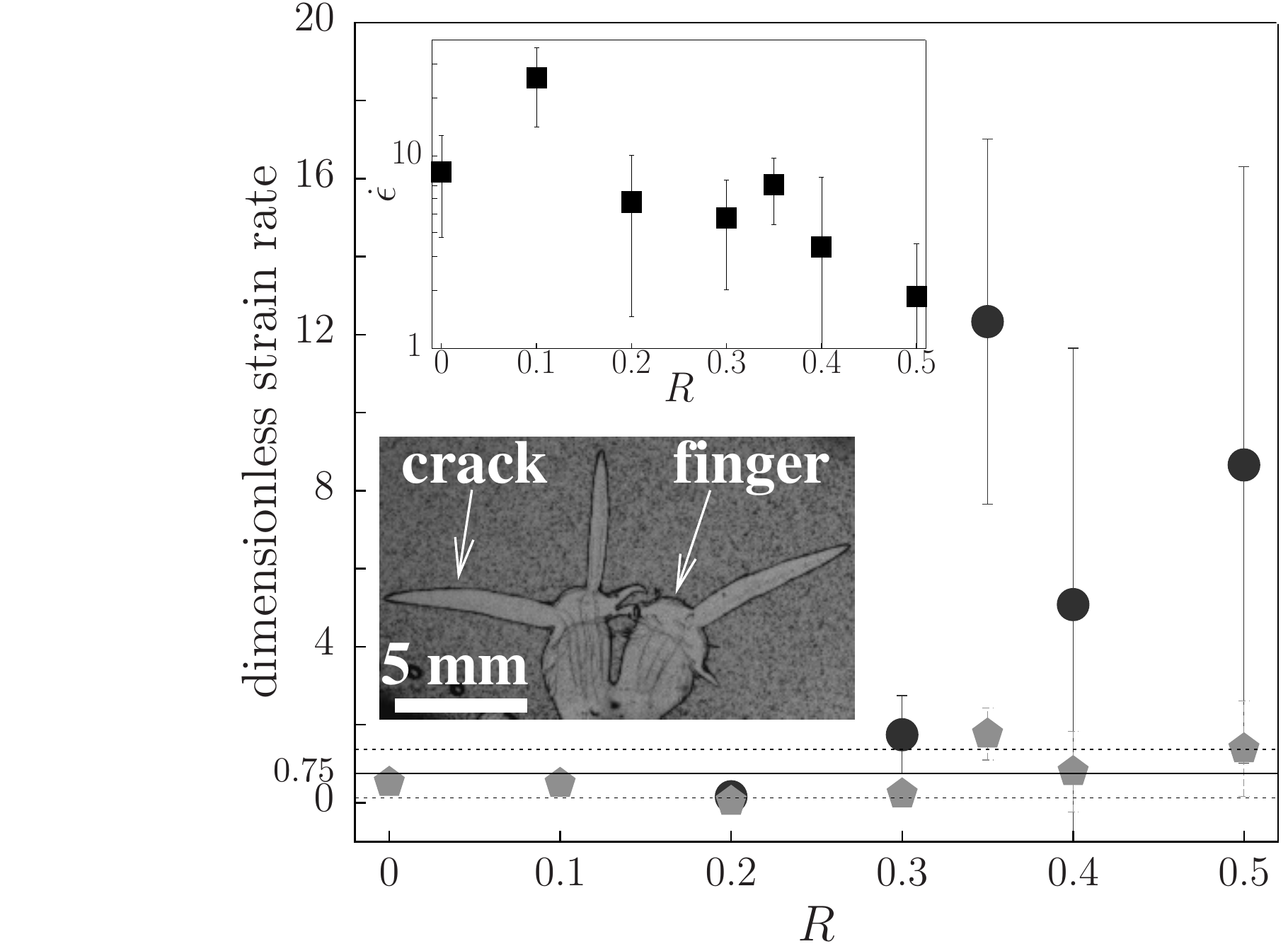}
\caption{Maximum strain rate at the F2F transition. Evolution with $R$ of {\color {myc} (inset)} the strain rate and {\color {myc}(main plot)} the strain rate normalized independently with the
shorter relaxation time ($\tau_{\rm{pol}}$, gray pentagons) and with the longer relaxation time ($\tau_{\rm{worm}}$, black circles).  {\color {myc} The continuous and dotted lines correspond to the mean value $\pm standard deviation$ of $\dot{\varepsilon}\tau_{\rm{pol}}$}. Image: Cracks
developing from fingers at the F2F transition (for a double network with $R=0.3$).}
\label{fig:F2F}
\end{figure}

The experimental set-up  has been described elsewhere~\cite{Foyart2013}. It consists in a standard radial Hele-Shaw cell (eventually sandwiched between crossed polarizers), where the sample is confined between two glass plates separated by $500 \, \mu$m thick Mylar spacers. Colza oil (zero-shear viscosity $60$ mPa.s) is injected at a constant volume rate, $Q$ [in the range $(0.01-40)$ ml/min], using a syringe pump through a hole drilled in one of the two plates, and pushes the gel. At low $Q$, a standard Saffmann-Taylor viscous fingering instability takes place. Above a threshold rate $Q_c$ that depends on the characteristic relaxation time(s) of the viscoelastic gel, a fingering to fracturing (F2F) transition occurs, and cracks propagate at a speed that depends on the injection rate. The visualization of the whole cell and of the oil-gel interface is achieved with a CMOS camera (Phantom v7), run at a $500$ Hz acquisition rate.
The first series of experiments is dedicated to the determination of the critical velocity for the F2F transition. To this aim, a fingering instability is first produced by injecting oil at low $Q$ ($0.01$ ml/min). An abrupt jump to a high $Q$ ($10$ ml/min) is then imposed, resulting in an acceleration of the finger propagation in the viscoelastic sample, before a transition occurs above a critical velocity $V^*$. A photograph of a F2F transition is provided in the inset of fig.~\ref{fig:F2F}. Those experiments were conducted with both single and double networks.
By contrast, in the second series of experiments we focus uniquely on a double network (with $R=0.5$), impose injection rates larger than $Q_c$, and image the crack propagation, with or without crossed polarizers.

We determine the maximal velocity of the tip of the finger $V^*$ at the F2F transition. For this velocity, the extension rate involved at the finger tip is approximated as $\dot{\varepsilon}\approx\frac{V^*}{\rho^*}$, where $\rho^*$ is the radius of curvature of the finger. We have experimentally determined $\rho^*$ and $V^*$ for samples with various $R$, quantifying at least five fingers per sample. $\rho$ is typically of the order of $1$ mm and $V^*$ of a few mm/s. {\color {myc} We measure that $\dot{\varepsilon}$ decreases as $R$ increases (inset of fig.~\ref{fig:F2F}).} We plot in fig.~\ref{fig:F2F} the evolution of $\dot{\varepsilon}\tau_{\rm{pol}}$ and $\dot{\varepsilon}\tau_{\rm{worm}}$ as a function of $R$ for all samples, {\color {myc} as these normalized quantities are the physically relevant parameters}. Here error bars correspond to the standard deviation of the measurements performed on the different fingers. For one-mode Maxwell fluids ($R=0$ and $R=0.1$), when the micelles are too short to be entangled, we find that the F2F transition occurs when the extensional rate becomes of the order of the characteristic relaxation time of the samples ($\dot{\varepsilon}\tau_{\rm{pol}}\approx 0.51 \pm 0.28$). Note that for similar samples made of oil droplets linked by the same telechelic polymers as the one used here~\cite{Foyart2013}, and for which the relaxation time was much larger than those of the samples investigated here ($\tau_{\rm{pol}}=0.7$ s), the same dependence holds ($\dot{\varepsilon}\tau_{\rm{pol}}\approx 0.7 $). We find here that this dependence also holds for the double networks ($\dot{\varepsilon}\tau_{\rm{pol}}\approx 0.75 \pm 0.62$, as averaged over all samples, {\color {myc} lines in} fig.~\ref{fig:F2F}). This suggests that crack nucleation is driven by the network of telechelic polymers  {\color {myc} as cracks occurs when the polymer network is strained at a too large rate to relax}. For the double networks, the characteristic time scale for the relaxation of the Maxwell is larger than that of the polymer ($\tau_{\rm{worm}} > \tau_{\rm{pol}}$) (fig.~\ref{fig:Rheo}). {\color {myc} Consequently, $\dot{\varepsilon}\tau_{\rm{worm}}\gg 1$}. This implies that, on the time scale for crack nucleation and propagation, the network of wormlike micelles does not have time to relax to its equilibrium structure.  Hence one may expect that the large stress involved at the crack tip leads to non relaxed rearrangements of the surfactant micelles. Because of the anisotropy of the micelles network, this may lead to local alignment of the micelles, as observed when semi-dilute wormlike micelles network are submitted to shear or elongations stresses~\cite{Berret2000,Handzy2004,Pathak2006,Ramos2011}.

\begin{figure}
\includegraphics[width=0.5\textwidth]{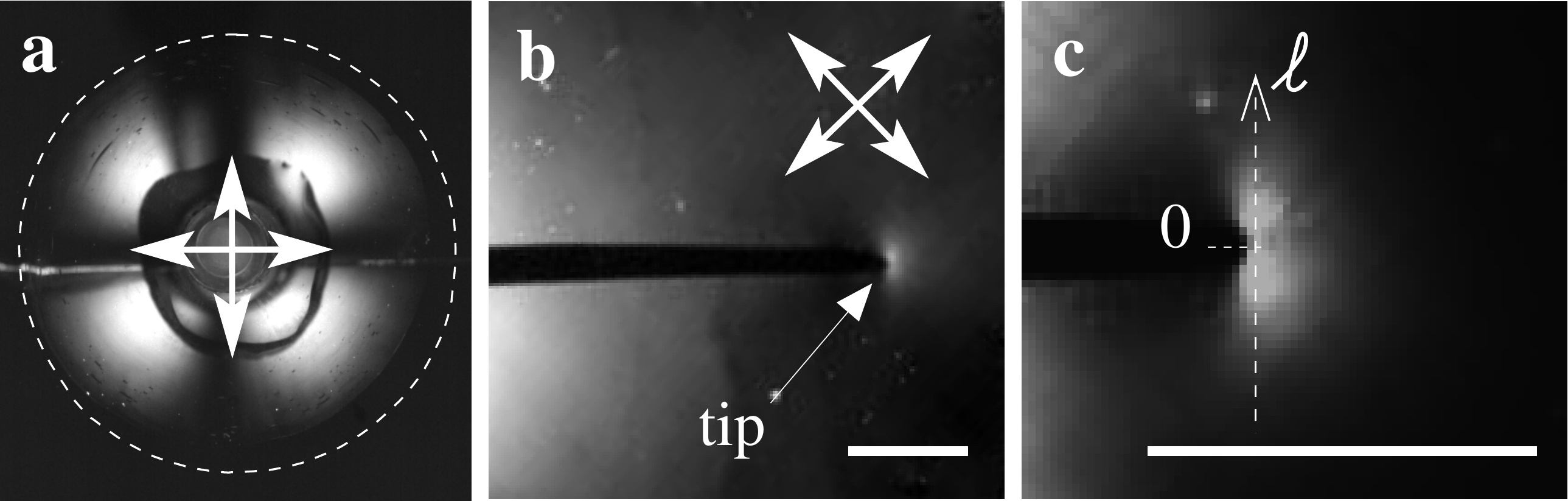}
\caption{Birefringence signal for a double network with $R=0.5$. (a) Snapshot taken before crack nucleation, while the oil is injected by the center of the cell. The dashed circle (true radius $8$ cm) marks the boundary of the region occupied by the gel. (b) Crack propagating radially. (c) Zoom of the crack tip shown in (b). The scale of the white bars is $1.5$ mm. The orientations of two crossed polarizers are indicated by the double arrows.}
\label{fig:ImagesBiref}
\end{figure}

\begin{figure}
 \includegraphics[width=0.5\textwidth]{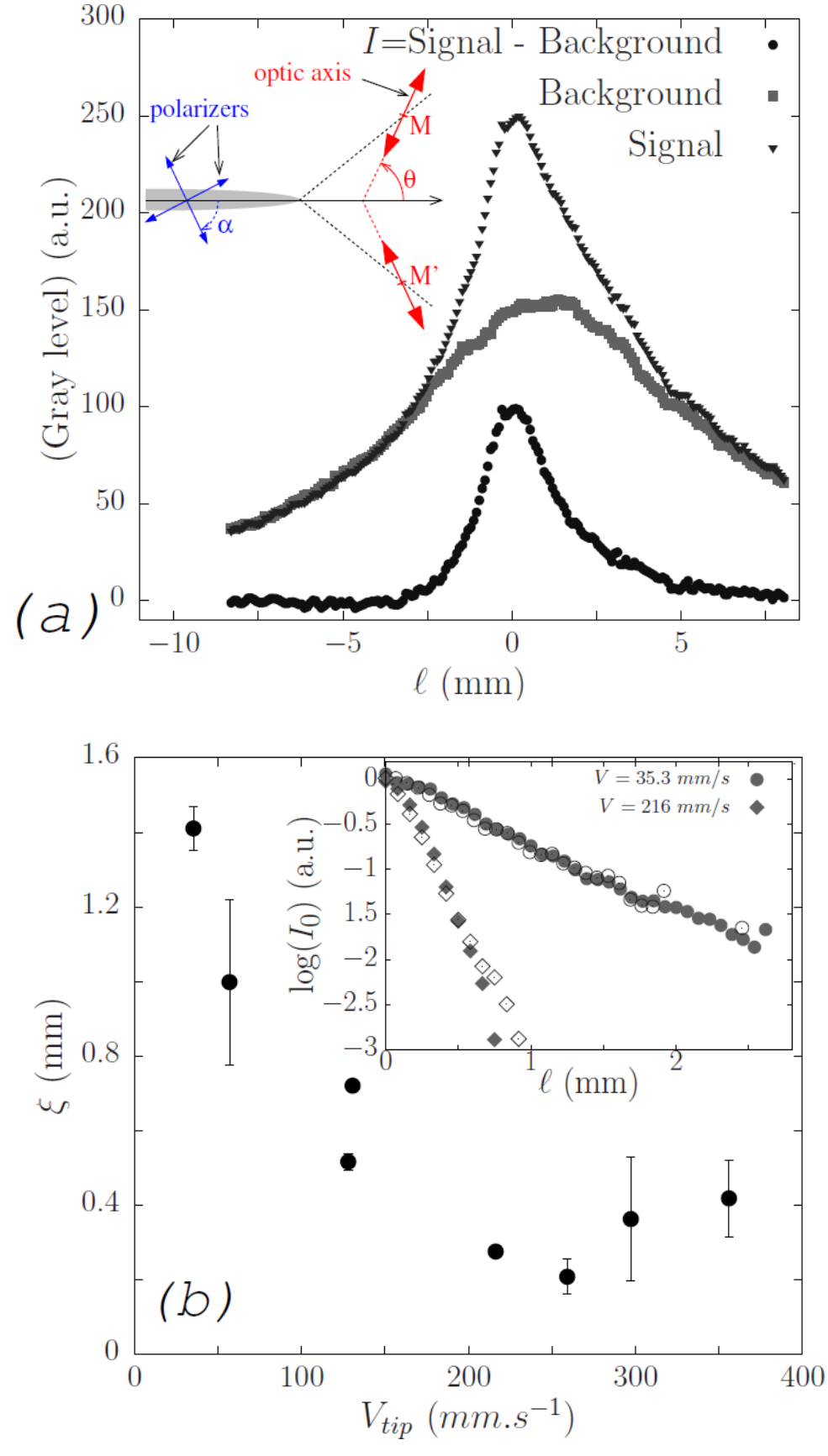}
\caption{Birefringence at the crack tip for a double network with $R=0.5$. (a) Birefringence intensity (black triangles) measured along the dashed line shown in Fig.~\ref{fig:ImagesBiref}c, background intensity (gray squares), and signal after background subtraction (black circles). Inset: Sketch for the calculation of the birefringence signal.
(b) Characteristic decay length of the birefringence as a function of the
speed of the crack tip. Inset:  corrected birefringence intensity in semi-log scale as obtained from Eqs. (1,2) for two tips velocities. Empty and filled symbols correspond to the two solutions Eqs. (1,2).}
\label{fig:Profiles}
\end{figure}

We focus here on a double network ($R=0.5$) that we image between crossed polarizers in order to quantify the birefringence related to the local alignment of the micelles, and we impose an injection rate $Q>Q_c$ to nucleate cracks. As shown in fig.~\ref{fig:ImagesBiref}a, the background displays some birefringence due to the alignment of the micelles at large scale because of the  mean radial flow. Its signature is a Maltese cross pattern with a signal that decreases with the distance from the injection point. Interestingly enough, as the crack propagates, one systematically measures an additional birefringence signal at the crack tip (fig.~\ref{fig:ImagesBiref}b,c). This indicates a preferential alignment of the micelles at the crack tip (although one cannot know experimentally whether the micelles are parallel or perpendicular to the direction of propagation of the crack~\cite{Trebbin2013}). Figure~\ref{fig:ImagesBiref}c shows that the birefringence pattern around the crack tip displays a peculiar butterfly shape, with a more intense and extended signal in the direction perpendicular to the crack propagation. We therefore quantify the birefringence signal in this direction (dotted line in fig.~\ref{fig:ImagesBiref}c). We show in fig.~\ref{fig:Profiles}a the intensity profile along this line, as a function of the distance $\ell$ from the tip.  We find a higher intensity at the crack tip that decays with the distance from the tip. We also show on the same plot the background taken just before the cracks enters the field of view.
The signal due to the crack tip is clearly much larger than the background signal around $\ell=0$ (close to the tip) and approaches that of the background far from the tip. The physically relevant signal, as obtained by subtracting the background from the measured profile, is also displayed in Fig.~\ref{fig:Profiles}a. This signal is generally asymmetric with respect to the crack tip position ($\ell=0$). The origin of this asymmetry results from the relative orientation of the direction of crack propagation and of the polarizer and analyzer and can be simply taken into account.
Let $M$ be a material point located at a distance $\ell$ perpendicularly to the crack tip, and assume the optic axis of the gel is inclined relatively to the propagation direction of the crack with the angle $\theta$ (see inset of Fig.~\ref{fig:Profiles}a). Let the angle of the analyzer and the polarizer with respect to the propagation direction be $\alpha \pm \frac{\pi}{2}$. The intensity transmitted through the sample slab and the polarizers, $I$, reads $I \approx I_0(h,\Delta n) \sin^2\left(2(\theta-\alpha) \right)$ (Eq. 1). Here,
$h$ is the thickness of the slab and $\Delta n$, the difference between the refractive indices of the ordinary and extraordinary rays, is assumed to be constant across the cell thickness. The intensity at $M'$, the point symmetric to $M$ with respect to the direction of propagation, is $I' \approx I_0(h,\Delta n) \sin^2\left(2(\theta+\alpha) \right)$ (Eq. 2).
For each pair ($I,I'$), i.e. for each distance $\ell$,  Eqs. (1,2) lead to two possible values for $I_0(h,\Delta n)$, without possible discrimination. We interestingly find that both solution for $I_0$ decay exponentially from the crack tip, thus defining each a finite length $\xi$. The lengthes corresponding to each of the two solutions for $I_0$ are found to be similar. The mean values of $\xi$ (error bars extremities correspond to the two individual solutions) are plotted in fig.~\ref{fig:Profiles}b as a function of the tip velocity $V_{\rm{tip}}$.
We find that $\xi$ is always on the order of the thickness of the cell ($0.5$ mm) but clearly decreases with the velocity of the tip, $V_{\rm{tip}}$, from $1.4$ mm to about $0.2$ mm, when the velocity increases from $35$ to $355$ m/s.

Within the linear elastic fracture mechanics theory, the long range decay of the strain field is expected to scale as $1/\sqrt{\ell}$ (for $\ell \gg b$, with $b$ the gap of the cell)~\cite{Irwin1957}. Therefore, the finite length scale measured here has a different origin and is due to micelle alignment in a well defined region. Birefringence in the vicinity of a crack in a viscoelastic wormlike micelles solution has been previously visualized~\cite{Gladden2007}. However no quantification was performed and the cracks were complex because occurring around a moving rod of square section.  A common argument would be that the relevant length scale for strain localization is the distance between neighboring junctions (the mesh size)~\cite{Erk2012}. Our findings are in sharp contrast with this statement as the length scale $\xi$ is much larger than the structural length scale (typically of a few nm). Our results are on the other hand in line with experimental results obtained for double gel networks consisting in interconnected densely crosslinked polyelectrolyte gel and loosely cross-linked  neutral polymer network where a permanent damage zone (thickness a few hundred micrometers, i.e. much larger than the mesh size of the network) has been visualized~\cite{Tanaka2008, Yu2009}. However, in Refs.~\cite{Tanaka2008, Yu2009}, the size of the damage zone was measured to {\color {myc} increase} with the speed of the crack, at odds with our findings, {\color {myc} showing that the nature of the network is important}. Notably, our experiments suggest that when the crack velocity increases, the gel appears less ductile as the characteristic length scale over which reorganization takes place decreases. This finding is in full agreement with experiments on two-dimensional dry foams which behave in a brittle-like manner as crack propagate rapidly and in a ductile-like manner at low speed~\cite{Arif2010}.

In conclusion we have investigated a ductile viscoelastic fluid and identified a rearrangement (process) zone around a crack tip whose size increases as the crack propagates slower, suggesting an increase ductility of the sample. We expect that our results would stimulate theoretical modeling.
{\color {myc} We believe that the method we have developed can be applied to other experimental configurations and systems, allowing ultimately rationalizing and controlling the dissipation mechanisms in fracture processes, by the design of the sample structure.}

\begin{acknowledgments}
We thank Ty Phou for polymer synthesis. This work has been supported by the French National Research Agency (ANR) under Contract No. ANR-2010-BLAN-0402-1 (F2F).
\end{acknowledgments}



\end{document}